\documentclass[pra,aps,showpacs,showkeys,twocolumn,twoside]{revtex4}

\usepackage{amsmath}
\usepackage{amssymb}
\usepackage{amscd}

\usepackage{latexsym}
\usepackage{epsfig}
\usepackage{graphics}
\newtheorem{defi}{Definition}
\newtheorem{lemma}[defi]{Lemma}
\newtheorem{thm}[defi]{Theorem}
\newtheorem{cor}[defi]{Corollary}
\newtheorem{rem}[defi]{Remark}

\newcommand{\qed}{\hfill $\Box$}
\newcommand{\tr}{{\operatorname{Tr}\,}}

\newcommand{\rank}{{\operatorname{rank}\,}}

\newcommand{\bra}[1]{{\langle{#1}|}}
\newcommand{\ket}[1]{{|{#1}\rangle}}
\newcommand{\ketbra}[1]{{\ket{#1}\!\bra{#1}}}

\newcommand{\fset}[1]{{\mathcal{#1}}}

\newcommand{\E}{{\mathbb{E}}}
\newcommand{\1}{{\openone}}

\newcommand{\strichr}{-\!\!\!-\!\!\!-\!\!}
\newcommand{\strichl}{\!\!-\!\!\!-\!\!\!-}
\newlength{\blank}
\newlength{\equalsign}
\newenvironment{beweis}[1][{\hspace{-\blank}}]{{\noindent\emph{Proof~{#1}.\ }}}{\hfill $\Box$\vskip 0.5\baselineskip}

\begin{document}

%\draft
%\twocolumn
%\narrowtext

\title{On the communication cost of entanglement transformations}
\author{Patrick Hayden}
\email{patrick@cs.caltech.edu}
\affiliation{Institute for Quantum Information, Caltech 107--81, Pasadena, CA 91125, USA}
\author{Andreas Winter}
\email{winter@cs.bris.ac.uk}
\affiliation{Department of Computer Science, University of Bristol,\\
Merchant Venturers Building, Woodland Road, Bristol BS8 1UB, United Kingdom}
\date{$31^{\rm st}$ May 2002}

% begun: February 1, 2002.

\begin{abstract}
  We study the amount of communication needed for two parties to transform
  some given joint pure state into another one, either exactly or with some
  fidelity.  Specifically, we present a method to lower bound this 
  communication cost even
  when the amount of entanglement does not increase.  Moreover, the bound
  applies even if the initial state is supplemented with unlimited 
  entanglement in the form of EPR
  (Einstein--Podolsky--Rosen) pairs and the communication is
  allowed to be quantum mechanical.
  \par
  We then apply the method to the determination of the communication cost
  of asymptotic entanglement concentration and dilution. While concentration
  is known to require no communication whatsoever, the best known protocol
  for dilution, discovered by Lo and Popescu 
  [Phys. Rev. Lett. 83(7):1459--1462, 1999], requires a number of bits
  to be exchanged which is of the order of the square root of the number of
  EPR pairs.  Here we prove a matching lower bound of the same asymptotic
  order, demonstrating the optimality of the Lo--Popescu protocol up to
  a constant factor and 
  establishing the existence of a fundamental asymmetry between the
  concentration and dilution tasks.
  \par
  We also discuss states for which the minimal communication cost
  is proportional to their entanglement, such as the states 
  recently introduced in the context of ``embezzling entanglement''
  [W. van Dam and P. Hayden, {\tt quant-ph/0201041}].
\end{abstract}



\maketitle

\section{Pure state entanglement transformations}
\label{sec:introduction}
The quantification of entanglement began with the study of the following 
question: assume that two parties, generically referred to as 
\emph{Alice} and \emph{Bob},
share $n$ copies of a bipartite pure state $\ket{\phi_{AB}}$ which by local
operations and classical communication (LOCC), they would like to convert
into a state that has high fidelity to $k$ copies of the target state 
$\ket{\psi_{AB}}$, with $k$ as large as possible. The basic question is then,
what is $\lim k/n$ as $n\rightarrow\infty$ and the fidelity goes to one?
\par
It turns out~\cite{bbps} that this optimal asymptotic ratio is equal to
$E(\phi)/E(\psi)$, where
$$E(\phi)=S(\tr_B\ketbra{\phi})=-\tr\bigl(\tr_B\ketbra{\phi}\log\tr_B\ketbra{\phi}\bigr)$$
is the von Neumann entropy of Alice's reduced state. For this reason, $E$ 
is often called the \emph{entropy of entanglement}.
One consequence of this result is that pure state entanglement can be 
interconverted \emph{asymptotically losslessly} between its different forms, 
justifying the introduction of the \emph{ebit} as a resource quantity, with
its ubiquitous ``incarnation'', the two--qubit EPR pair state, which,
up to a local change of basis, can be written as
$$\ket{\phi_2^+}=\frac{1}{\sqrt{2}}\bigl(\ket{00}+\ket{11}\bigr).$$
For the sake of quantifying entanglement, however, not only local actions by 
Alice and Bob but \emph{classical communication} was considered unlimited.
It is precisely these communication requirements that we study in the
present paper, in which we follow Lo's~\cite{lo:ccc} suggestion to study
the communication complexity of distributed quantumn information processing.
\par
This goal notwithstanding, our point of departure will not be the theory 
of asymptotically faithful transformations but, rather, its finite (and 
more refined) variant of transformations from $\ket{\phi_{AB}}$ to 
$\ket{\psi_{AB}}$ up to fidelity $1-\epsilon$, as laid out 
in~\cite{vjn:enttrans}, building on previous work~\cite{nielsen},
\cite{hardy}, \cite{lo:popescu:beyond} for the zero--error case.
\par
Up to local unitaries, pure entangled states are uniquely defined by the
spectrum of their reduced states (either at Alice's or Bob's side),
the eigenvalues known as the \emph{Schmidt coefficients} $\lambda_j$.
Indeed, it is possible to choose bases in the entangled system such that
$$\ket{\phi_{AB}}=\sum_j \sqrt{\lambda_j}\ket{i}_A\otimes\ket{i}_B.$$
The theory relates the feasibility of such an LOCC transformation
to the \emph{majorisation order} of the Schmidt coefficients
$(\lambda)$ of $\ket{\phi}$ and $(\mu)$ of $\ket{\psi}$, both vectors
arranged in nonincreasing order:
$$\ket{\phi}\stackrel{\text{LOCC}}{\strichr\!\strichr\!\longrightarrow}\ket{\psi}
    \quad\text{iff}\quad
  (\lambda) \prec (\mu),$$
where $(\lambda) \prec (\mu)$ is defined to mean
\begin{equation*}
  \forall k\quad \sum_{j=1}^k \lambda_j \leq \sum_{j=1}^k \mu_j,
\end{equation*}
which can be shown to be equivalent to the existence of a doubly
stochastic matrix $M$ such that~$(\lambda)=M\cdot(\mu)$.
By the results of~\cite{hardy} and \cite{jensen:schack}, any such allowed 
transformation can always be achieved by one--way communication, say from
Alice to Bob, of $2\log\rank\tr_B\ketbra{\phi}$ classical bits.
\par
The organisation of the paper is as follows. In section~\ref{sec:lower} 
we will explain the mathematical model of approximate
pure state transformations and derive our main result, a lower bound on
the communication cost of state transformations which holds even
if the initial state is supplemented by an unlimited
number of EPR pairs, and even if the communication is quantum mechanical.
To our knowledge this is the first \emph{quantitative} statement
of its kind. (The need for some nonzero amount of communication in certain
transformations was pointed out in~\cite{lo:popescu:beyond}.)
We then apply the result in section~\ref{sec:dilution} to the asymptotic
transformations mentioned in the introduction, proving a lower bound of
$\Omega(\sqrt{n})$ on the communication necessary for 
entanglement dilution, which, up to a constant factor, matches 
the $O(\sqrt{n})$ construction of
Lo and Popescu~\cite{lo:popescu:sqrt} for this task.
In section~\ref{sec:large:cost} we analyse a class of states that require
for their creation from EPR pairs communication of the same order
as their entanglement, before ending with a discussion of some open 
problems.

\section{A lower bound on the communication cost}
\label{sec:lower}
Assume that initally Alice and Bob share the state $\ket{\phi}$,
then execute several rounds of local actions and classical communication,
and finally end up with some joint state $\widetilde{\rho}$ that has high
fidelity to $\ket{\psi}$.
Allowing the use of \emph{quantum bits} to communicate, we give Alice and
Bob even more power, thereby potentially reducing the communication cost, 
while at the
same time simplifying the appearance of the protocol: because each of the
local actions can be implemented using ancillae and unitary transformations,
the whole process can be reduced to a series of exchanges of quantum systems
of certain dimensions $d_i$ between Alice and Bob,
with a final tracing out (discarding) of part of Alice's and part of Bob's 
system. The total communication cost of such a procedure is just
$C=\sum_{i=1}^N \log d_i$ qubits.
\par\medskip
\begin{figure}[ht]
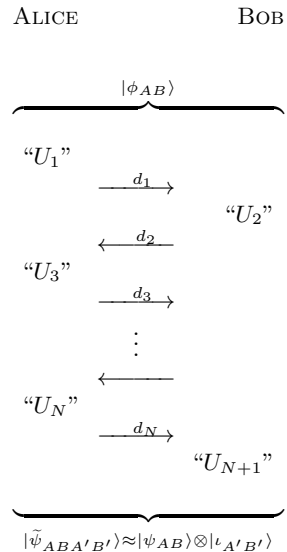

  \label{fig:protocol}
  \centerline{\sc Alice \phantom{=======} Bob}
  \begin{equation*}
    \underbrace{\overbrace{%
    \begin{array}{lcr}
      & & \\
      \text{``}U_1\text{''} &                                                     &                 \\
                            & \strichr^{d_1}\!\!\!\!\!\!\longrightarrow           &                 \\
                            &                                               & \text{``}U_2\text{''} \\
                            & \longleftarrow^{d_2}\!\!\!\!\!\!\strichl            &                 \\
      \text{``}U_3\text{''} &                                                     &                 \\
                            & \strichr^{d_3}\!\!\!\!\!\!\longrightarrow           &                 \\
                            &                    \vdots                           &                 \\
                            & \longleftarrow^{\phantom{d_1}}\!\!\!\!\!\!\strichl  &                 \\
      \text{``}U_N\text{''} &                                                     &                 \\
                            & \strichr^{d_N\!}\!\!\!\!\!\!\!\longrightarrow       &                 \\
                            &                                           & \text{``}U_{N+1}\text{''} \\
      & & \\
    \end{array}
    }^{\ket{\phi_{AB}}}
    }_{\ket{\widetilde{\psi}_{ABA'B'}}\approx\ket{\psi_{AB}}\otimes\ket{\iota_{A'B'}}}
  \end{equation*}
  \caption{In round $i$ Alice (Bob) performs some unitary $U_i$ on her (his) system,
    which separates into a residual system and a $d_i$--dimensional system that
    is sent to Bob (Alice). 
    In the last, $N^{\rm th}$, round, the receiver of the message may perform
    a unitary on his/her system, and then Alice and Bob trace out
    subsystems $A'$ and $B'$.}
\end{figure}
\par
The idea of the lower bound is very simple, and is explained most 
straightforwardly for \emph{exact} state transformations, 
when $\widetilde{\rho}=\ketbra{\psi}$.
During the process of transformation we monitor a certain quantity
$\Delta$ associated to Alice's reduced density operator,
showing that, for each qubit communicated, it can only increase
by a constant, and then observe that the final partial trace never
increases $\Delta$ at all. The difference between the initial and
the final $\Delta$ then provides a lower bound on the communication.
\par
Specifically we shall consider, for $\rho=\tr_B\ketbra{\phi}$,
\begin{equation}
  \label{eq:delta:0}
  \Delta(\rho):=S_0(\rho)-S_\infty(\rho),
\end{equation}
where $S_\alpha$ are the \emph{R\'{e}nyi entropies}~\cite{renyi} of order $\alpha$:
$$S_\alpha(\rho):=\frac{1}{1-\alpha}\log\tr(\rho^\alpha).$$
For $\alpha=0,1,\infty$ the R\'{e}nyi entropies are defined by continuous 
extension, with resulting formulas
\begin{align*}
  S_0(\rho)      &= \log\rank\rho,       \\
  S_1(\rho)      &= -\tr(\rho\log\rho),  \\
  S_\infty(\rho) &= -\log\|\rho\|_\infty,
\end{align*}
where $\|\cdot\|_\infty$ is the supremum norm: for selfadjoint operators it
is the largest absolute value of an eigenvalue.
(Throughout the paper, $\log$ and $\exp$ are understood to be base $2$.)
Note that $\Delta(\rho)\geq 0$ since $S_\alpha(\rho)$ is nonincreasing
in $\alpha$~\cite{renyi}, or by inspection of the definition.
Furthermore, if all the non--zero eigenvalues of $\rho$ are
the same then $S_0(\rho)=S_\infty(\rho)$ so that $\Delta(\rho)=0$.
Otherwise, $\Delta(\rho)$ will be strictly greater than zero.  Therefore, 
$\Delta(\rho)$ can be interpreted as a measure of the variation in the 
eigenvalues of $\rho$.
\par
The key observation is that, in communication round $i$, the R\'{e}nyi
entropy of Alice's reduced state, whose spectrum characterises the
entanglement, cannot change too much.  To see this, we assume without loss 
of generality that it is Alice's turn to perform a unitary, rotating her
reduced state to $\rho_{AA'}$.  This step, obviously, does not change
the R\'{e}nyi entropy at all.  Next, she gives Bob the $d_i$--dimensional 
system $A'$, leaving her with the new reduced state $\rho_A$, 
for which we have the relation (see~\cite{comm:compl})
\begin{equation}
  \label{eq:renyi:change}
  S_\alpha(\rho_{AA'})-\log d_i \leq S_\alpha(\rho_A) \leq S_\alpha(\rho_{AA'})+\log d_i,
\end{equation}
which implies (inserting $\alpha=0,\infty$)
\begin{equation}
  \label{eq:Delta:change}
  \Delta(\rho_A) \leq \Delta(\rho_{AA'}) + 2\log d_i.
\end{equation}
Thus, the quantity $\Delta$ can increase (or decrease) by at most $2\log d_i$
in step $i$.
After the last round of communication has taken place, the joint state is
$\ket{\widetilde{\psi}_{ABA'B'}}=\ket{\psi_{AB}}\otimes\ket{\iota_{A'B'}}$.
(Note that if this were not a product state, $\widetilde{\rho}$ would
necessarily not be pure.) Hence, by induction over the number of rounds,
summing over the the eqs.~(\ref{eq:Delta:change}) yields
\begin{equation}
  \label{eq:Delta:C}
  \Delta\bigl(\tr_{BB'}\ketbra{\widetilde{\psi}_{ABA'B'}}\bigr)
                         \leq \Delta\bigl(\tr_B\ketbra{\phi}\bigr) + 2C.
\end{equation}
The effect on $\Delta$ of the final partial trace over the primed system
is easy to understand: because the R\'{e}nyi entropies are additive 
under tensor products, i.e.
$$S_\alpha(\rho\otimes\sigma)=S_\alpha(\rho)+S_\alpha(\sigma),$$
we obtain
\begin{equation}\begin{split}
  \label{eq:factor:off}
  \Delta&\bigl(\tr_B\ketbra{\psi}\otimes\tr_{B'}\ketbra{\iota}\bigr)  \\
        &\phantom{======}
         =\Delta\bigl(\tr_B\ketbra{\psi}\bigr)+\Delta\bigl(\tr_{B'}\ketbra{\iota}\bigr),
\end{split}\end{equation}
and the rightmost term is nonnegative.
This proves
\begin{thm}
  \label{thm:bound:0}
  A (deterministic) pure state transformation of $\ket{\phi_{AB}}$ into
  $\ket{\psi_{AB}}$ requires at least
  $$C \geq \frac{1}{2}\Bigl( \Delta\bigl(\tr_B\ketbra{\psi}\bigr)
                            -\Delta\bigl(\tr_B\ketbra{\phi}\bigr) \Bigr)$$
  bits of communication, even if quantum communication is allowed.
  \qed
\end{thm}
We note that in \cite{comm:compl} the analogous theorem for the 
bare R\'{e}nyi entropies
$S_\alpha$ was used to prove bounds on the communication required to
perform entanglement transformations in an approximate setting.  There,
changes in $S_\alpha$ reflected changes in the amount of entanglement
present in the system.  The advantage of using $\Delta$ is precisely that
it does not measure entanglement but, rather, variation in the Schmidt 
coefficients.
\begin{rem}
  \label{rem:alpha:beta}
  Obviously, a similar result holds for
  $$\Delta^{\alpha\beta}(\rho):=S_\alpha(\rho)-S_\beta(\rho),$$
  with arbitary $0 \leq \alpha < \beta \leq \infty$.
  Even though $\Delta^{\alpha\beta}(\rho)\leq\Delta(\rho)$, for some
  $\alpha$ and $\beta$ the \emph{increase} of
  the former quantity in an entanglement transformation may exceed the
  increase of the latter.
\end{rem}
\par
\begin{rem}
  \label{rem:epr2epr}
  As an example of a nontrivial consequence of theorem~\ref{thm:bound:0}
  we may observe that it puts severe restrictions on the entanglement
  transformations possible \emph{without any communication}:
  none of the $\Delta^{\alpha\beta}$ must increase.
  \par
  For example, from a maximally entangled state only other maximally
  entangled states (with possibly smaller Schmidt rank) may be obtained.
  If the Schmidt rank of the target divides that of the initial state
  this is clearly possible, while inspection of eq.~(\ref{eq:factor:off})
  shows that this is also necessary.
\end{rem}
\par
For the case of high--fidelity transformations this approach turns out to
be too simple: neither $S_0$ nor $S_\infty$ can be well controlled if
we switch from a state to one close by.  For example,
for the dilution task, which consists of the creation of 
$\bigl(\alpha\ket{00}+\beta\ket{11}\bigr)^{\otimes n}$
from EPR pairs, theorem~\ref{thm:bound:0}
implies a lower bound of $\Omega(n)$, while we know
from~\cite{lo:popescu:sqrt} that arbitrarily high fidelity can be achieved
with $O\bigl(\sqrt{n}\bigr)$ bits of communication.
\par
Instead, we invent robust versions of $S_0$, $S_\infty$ and $\Delta$:
let the eigenvalues of $\rho$ be denoted $r_j$ 
and then define, for $0 \leq \epsilon < 1$,
\begin{align}
  \label{eq:S:0:epsilon}
  S_{0,\epsilon}(\rho)      &:=  \log\min\!\left\{|J| : \sum_{j\in J} r_j \!\geq 1-\epsilon\right\}\!,\\
  \label{eq:S:infty:epsilon}
  S_{\infty,\epsilon}(\rho) &:= -\log\min\!\left\{\max_{j\in J} r_j
                                                   : \sum_{j\in J} r_j \!\geq 1-\epsilon\right\}\!,  \\
  \label{eq:Delta:epsilon}
  \Delta_{\epsilon}(\rho)  &:=  \log\min\!\left\{|J|\left(\max_{j\in J} r_j\right)
                                                   : \sum_{j\in J} r_j \!\geq  1-\epsilon\right\}\!,
\end{align}
all the minimisations are understood to be over subsets $J$ of the eigenvalue 
indices $j$.
Note that
\begin{equation}
  \label{eq:piecemeal}
  \Delta_{\epsilon}(\rho) \geq S_{0,\epsilon}(\rho)-S_{\infty,\epsilon}(\rho),
\end{equation}
with equality generally only if $\epsilon=0$, in which case these quantities reduce
to the above $S_0$, $S_\infty$ and $\Delta$.
\begin{rem}
  \label{rem:alternative}
  Note that $\Delta_\epsilon$ has the following ``high--fidelity'' relation
  to $\Delta_0$:
  $$\Delta_\epsilon(\rho)=\min\left\{\Delta_0\left(P\rho P\right):
                                                   \tr(\rho P)\geq 1-\epsilon\right\},$$
  where the minimisation is over all projections $P$ commuting with $\rho$,
  extending the definition of $\Delta_0$ to sub--normalised density operators.
  The operators $P\rho P$ can be interpreted as post--measurement states
  after the event ``$P$'' has occurred, normalised to the event probability.
  \par
  More generally, we could allow any $0\leq B\leq\1$ in the above
  minimisation, such that $\tr(\rho B)\geq 1-\epsilon$, and substituting the
  post--measurement states $\sqrt{B}\rho\sqrt{B}$.
  (By a result in~\cite{winter:qstrong} this operator has high fidelity to the state $\rho$.)
  It is easy to see that the resulting quantity is within a
  distance of $\log(1-\epsilon)$ from $\Delta_\epsilon$.
\end{rem}
\par
We now prove a few lemmas which will together comprise our method of estimating
the communication cost, by providing the tools to estimate $\Delta_\epsilon$ 
for the appropriate reduced states.
We begin with the simple observation that for
all states $\rho$ and $\epsilon'<\epsilon<1$:
\begin{align}
  \label{eq:lower}
  \Delta_\epsilon(\rho) &\geq \log(1-\epsilon),        \\
  \label{eq:monotone}
  \Delta_\epsilon(\rho) &\leq \Delta_{\epsilon'}(\rho).
\end{align}
\begin{lemma}
  \label{lemma:fidelity}
  If for two states $\rho$ and $\sigma$, $\|\rho-\sigma\|_1\leq\epsilon$, then
  $$\Delta_0(\rho) \geq \Delta_{\sqrt{\epsilon}}(\sigma)+\log\left(1-\sqrt{\epsilon}\right).$$
  (Where $\|\cdot\|_1$ is the trace norm, for selfadjoint operators given by the
  sum of the absolute values of all eigenvalues, counting multiplicities.)
\end{lemma}
\begin{beweis}
  To begin, denote the eigenvalue lists of $\rho$ and $\sigma$
  by $(r)$ and $(s)$, respectively, in nonincreasing order.
  Then, because (see~\cite{nielsen:chuang})
  $$\|(r)-(s)\|_1 \leq \|\rho-\sigma\|_1 \leq \epsilon,$$
  we may concentrate on the eigenvalues only.
  Define, for $\delta=\sqrt{\epsilon}$,
  $$J:=\bigl\{j: (1-\delta)s_j \leq r_j \leq (1+\delta)s_j\bigr\}.$$
  Then, for the complement $J^{\rm c}$ of $J$,
  \begin{equation*}
    \delta s\left(J^{\rm c}\right) =    \sum_{j\not\in J} \delta s_j
                                   \leq \sum_{j\not\in J} |r_j-s_j|
                                   \leq \epsilon,
  \end{equation*}
  implying
  $$\sum_{j\in J} s_j \geq 1-\sqrt{\epsilon}.$$
  We may clearly assume that $s$ is nonzero on $J$, otherwise shrinking
  $J$ without affecting the last inequality.
  \par
  Thus, by the definition of $\Delta_{\sqrt{\epsilon}}$,
  $$\log\left(|J| \max_{j\in J} s_j\right) \geq \Delta_{\sqrt{\epsilon}}(\sigma).$$
  On the other hand, by the definition of $J$, 
  $$j\in J \Longrightarrow r_j\neq 0,$$
  which implies that $\rank\rho \geq |J|$. Similarly,
  $$j\in J \Longrightarrow r_j \geq \left(1-\sqrt{\epsilon}\right)s_j,$$
  implies $\max_j r_j \geq \left(1-\sqrt{\epsilon}\right) \max_{j\in J} s_j$.
  Comparing the last two observations to the definition of 
  $\Delta_0(\rho)$ finishes the proof of the claim.
\end{beweis}
\begin{lemma}
  \label{lemma:product}
  For any two states $\tau$ and $\omega$, and $\epsilon<1$,
  $$\Delta_\epsilon(\tau\otimes\omega) \geq \Delta_{\sqrt{\epsilon}}(\tau)
                                           +\log\left(1-\sqrt{\epsilon}\right).$$
\end{lemma}
\begin{beweis}
  Denote the eigenvalues of $\tau$ and $\omega$ by $t_i$ and $w_k$, respectively.
  Let $J$ be a set of indices $i$ and $k$ such that
  $$\Delta_\epsilon(\tau\otimes\omega)=\log\bigl(|J|\max_{ik\in J} t_i w_k\bigr),$$
  and 
  \begin{equation}
    \label{eq:product:constraint}
    (t\otimes w)(J)=\sum_{ik\in J} t_i w_k \geq 1-\epsilon.
  \end{equation}
  We shall be interested, for certain $k$, in the \emph{sections}
  $$S_k:=\bigl\{i:ik\in J\bigr\}$$
  \emph{of $J$ along $k$}, in particular in the set
  $$K:=\left\{k:t(S_k)=\sum_{i\in S_k} t_i \geq 1-\sqrt{\epsilon}\right\}.$$
  It follows from the definition of $K$ and the constraint of
  eq. (\ref{eq:product:constraint}) that
  \begin{equation}
    \label{eq:K:prob}
    w(K)=\sum_{k\in K} w_k \geq 1-\sqrt{\epsilon}.
  \end{equation}
  The proof is a standard Markov inequality argument:
  observe that we can rewrite eq.~(\ref{eq:product:constraint})
  using the sections:
  $$1-\epsilon \leq (t\otimes w)(J) = \sum_k w_k t(S_k).$$
  Now the right hand side is a probability average over the values $t(S_k)$,
  taken with probability $w_k$. We decompose the sum into two contributions
  which we estimate separately:
  \begin{equation*}\begin{split}
    1-\epsilon &\leq \sum_{k\in K} w_k t(S_k) + \sum_{k\not\in K} w_k t(S_k) \\
               &\leq w(K) + \bigl(1-w(K)\bigr)\bigl(1-\sqrt{\epsilon}\bigr).
  \end{split}\end{equation*}
  Hence $\bigl(1-w(K)\bigr)\sqrt{\epsilon}\leq\epsilon$, which is our claim.
  \par
  Now define
  $$J':=\bigcup_{k\in K} S_k\times\{k\},$$
  and successively estimate
  \begin{equation*}\begin{split}
    |J|\max_{ik\in J}(t_i w_k) &\geq |J'|\max_{ik\in J'}(t_i w_k)          \\
                               &=    \sum_{l\in K} |S_l|\max_{ik\in J'}(t_i w_k)  \\
                               &\geq \sum_{k\in K} |S_k|\max_{i\in S_k}(t_i w_k)  \\
                               &=    \sum_{k\in K} w_k \left( |S_k|\max_{i\in S_k}t_i \right) \\
                      &\geq \sum_{k\in K} w_k \exp\left(\Delta_{\sqrt{\epsilon}}(\tau)\right) \\
                      &\geq \left(1-\sqrt{\epsilon}\right)
                              \exp\left(\Delta_{\sqrt{\epsilon}}(\tau)\right),
  \end{split}\end{equation*}
  the second last line because of $t(S_k)\geq 1-\sqrt{\epsilon}$, the last
  line by eq.~(\ref{eq:K:prob}), which proves the lemma.
\end{beweis}
\begin{rem}
  \label{rem:quasi:superadd}
  We do not know if a symmetric version of this lemma holds, with an additional
  term to the right analogous to the one for $\tau$:
  $$\Delta_\epsilon(\tau\otimes\omega) \stackrel{{\rm ?}}{\geq}
                         (1-\epsilon')\Delta_{\epsilon'}(\tau)
                        +(1-\epsilon')\Delta_{\epsilon'}(\omega)+\epsilon'',$$
  with $\epsilon'$, $\epsilon''$ functions of $\epsilon$ which vanish for
  $\epsilon\rightarrow 0$.
  \par
  This would constitute a form of ``quasi--additivity'' for $\Delta$, since
  the validity of the analogous reverse inequality
  $$\Delta_{2\epsilon}(\tau\otimes\omega) \leq \Delta_\epsilon(\tau)
                                                +\Delta_\epsilon(\omega)$$
  is quite easy to see. While it may not be useful to improve on
  our present results, confirmation of the ``quasi-additivity'' would 
  be of conceptual interest.
\end{rem}
We are now ready to state our central result, which applies whenever the
output state has high Uhlmann fidelity
$F(\sigma,\omega) = \left( {\rm Tr} \sqrt{\sigma^{1/2} \omega \sigma^{1/2}}\right)^2$
\cite{jozsa,uhlmann} with the 
desired state, even if the output is mixed:
\begin{thm}
  \label{thm:bound}
  Consider a state transformation protocol that takes $\ket{\phi_{AB}}$
  to $\ket{\psi_{AB}}$ with fidelity $1-\epsilon$, exchanging a total
  of $C$ qubits in the process. Then, with $\delta=\sqrt[8]{4\epsilon}$,
  \begin{equation*}
    2C \geq \Delta_{\delta}\bigl(\tr_B\ketbra{\psi}\bigr)
           -\Delta_0\bigl(\tr_B\ketbra{\phi}\bigr)+2\log(1-\delta).
  \end{equation*}
\end{thm}
\begin{beweis}
  Like in the zero--error case, we follow the increase of $\Delta_0$ over the
  course of the protocol: after the last communication has taken place,
  the joint state is $\ket{\widetilde{\psi}_{ABA'B'}}$, and we have
  (compare eq.~(\ref{eq:Delta:C}))
  $$2C \geq  \Delta_0\bigl(\widetilde{\psi}_{AA'}\bigr)
            -\Delta_0\bigl(\tr_B\ketbra{\phi}\bigr),$$
  where $\widetilde{\psi}_{AA'}=\tr_{BB'}\ketbra{\widetilde{\psi}}$.
  \par
  Now, since $\tr_{A'B'}\ketbra{\widetilde{\psi}}$ has fidelity
  $1-\epsilon$ to $\ket{\psi_{AB}}$, we can choose a pure state
  $\ket{\iota_{A'B'}}$ such that
  $$F\bigl( \ket{\widetilde{\psi}_{ABA'B'}},
            \ket{\psi_{AB}}\otimes\ket{\iota_{A'B'}} \bigr)\geq 1-\epsilon.$$
  Introducing $\psi_A=\tr_B\ketbra{\psi}$ and
  $\iota_{A'}=\tr_{B'}\ketbra{\iota}$, we infer, from the monotonicity of
  the fidelity, that
  $$F\bigl( \widetilde{\psi}_{AA'}, \psi_A\otimes\iota_{A'} \bigr) \geq 1-\epsilon,$$
  from which it follows by standard inequalities~\cite{nielsen:chuang} that
  $$\| \widetilde{\psi}_{AA'}-\psi_A\otimes\iota_{A'} \|_1 \leq \sqrt{4\epsilon}.$$
  Now we can use lemma~\ref{lemma:fidelity}
  to lower bound $\Delta_0\bigl(\widetilde{\psi}_{AA'}\bigr)$
  in terms of $\Delta_{\sqrt[4]{4\epsilon}}\bigl(\psi_A\otimes\iota_{A'}\bigr)$,
  which is bounded in turn, using lemma~\ref{lemma:product}, by
  $\Delta_{\sqrt[8]{4\epsilon}}\bigl(\psi_A)$, which proves the theorem.
\end{beweis}
\par
Using the additivity of the R\'{e}nyi entropies,
and that $S_\alpha\bigl(\frac{1}{2}\1\bigr)=1$ for all $\alpha$,
we observe that
$$\Delta_0\left(\rho\otimes\frac{1}{2}\1\right)=\Delta_0(\rho).$$
This implies
\begin{cor}
  \label{cor:unlimited:EPRpairs}
  The lower bound on $C$ of theorem~\ref{thm:bound} continues to hold
  even if the starting state $\ket{\phi_{AB}}$ is supplemented by 
  unlimited numbers of EPR pairs.
  \qed
\end{cor}
Now suppose that $\ket{\phi_{AB}}$ can be converted
into a high-fidelity copy of $\ket{\psi_{AB}}$ using an LOCC protocol
in which only $C$ bits are exchanged between Alice and Bob.
By consuming EPR pairs for superdense coding \cite{bennett:wiesner}, 
this protocol
can be converted into a protocol requiring only $C/2$ qubits of communication.
Since the lower bound of the corollary applies to the modified protocol, 
we conclude that for classical communication our bound can be improved 
by a factor of two.
\begin{cor}
  \label{cor:classical}
  If the state transformation $\ket{\phi_{AB}}$ to $\ket{\psi_{AB}}$ can
  be accomplished with fidelity $1-\epsilon$ by exchanging a total of
  $C$ classical bits then, with $\delta= \sqrt[8]{4\epsilon}$,
  \begin{equation*}
     C \geq \Delta_{\delta}\bigl(\tr_B\ketbra{\psi}\bigr)
           -\Delta_0\bigl(\tr_B\ketbra{\phi}\bigr)+2\log(1-\delta).
  \end{equation*}
  \qed
\end{cor}
\begin{rem}
  \label{rem:robust}
  Sometimes, direct application of these results can give an overly 
  conservative lower bound
  because $\Delta_0\left(\tr_B\ketbra{\phi}\right)$ can be much larger than
  the corresponding $\Delta_\epsilon\left(\tr_B\ketbra{\phi}\right)$.
  \par
  Here we note that a lower bound on $C$ in terms of $\Delta_\epsilon$
  of both the initial and the final state exists: simply observe that
  changing the initial state $\ket{\phi}$ into some state $\ket{\phi'}$
  with fidelity $1-\epsilon_0$, the protocol results in a state $\rho'$
  that has fidelity $1-\epsilon_0$ to $\rho$ (because the fidelity does not
  decrease under completely positive trace preserving maps), which
  in turn has fidelity $1-\epsilon$ to $\ket{\psi}$.
  By a result of~\cite{bfjs} this implies that the transformation from
  $\ket{\phi'}$ to $\ket{\psi}$ has fidelity $1-\epsilon'$, with some universal
  function $\epsilon'$ of $\epsilon$ and $\epsilon_0$.
  We may then apply theorem~\ref{thm:bound} to this transformation.
\end{rem}
\begin{rem}
  \label{rem:alpha:beta:robust}
  Of course one can also define a robust version of our previous
  $\Delta^{\alpha\beta}$ (see remark~\ref{rem:alpha:beta}):
  $$\Delta_\epsilon^{\alpha\beta}(\rho):=
        \min\left\{ \frac{\log\left(\sum_{j\in J} r_j^\alpha\right)}{1-\alpha}
                   -\frac{\log\left(\sum_{j\in J} r_j^\beta\right)}{1-\beta} \right\},$$
  again with minimisation over all subsets of indices $J$ such that
  $\sum_{j\in J} r_j \geq 1-\epsilon$.
  Unsurprisingly, a variant of theorem~\ref{thm:bound} also holds for this
  quantity:
  \par
  Consider an entanglement transformation from $\ket{\phi}$ to
  $\ket{\psi}$ with fidelity $1-\epsilon$ and a total communication cost of
  $C$ qubits. Then, for $0\leq\alpha<1<\beta\leq\infty$,
  $$2C\geq \Delta_{\delta}^{\alpha\beta}\bigl(\tr_B\ketbra{\psi}\bigr)
                 -\Delta_0^{\alpha\beta}\bigl(\tr_B\ketbra{\phi}\bigr)+\delta',$$
  with $\delta=\sqrt[8]{4\epsilon}$ and
  $\delta'=\left(\frac{2\alpha}{1-\alpha}+\frac{2\beta}{\beta-1}\right)
             \log\bigl(1-\sqrt{\delta}\bigr)$.
  \par
  The proof is slightly more cumbersome version of the proof for the 
  $\Delta_\delta=\Delta_\delta^{0\infty}$ case.
\end{rem}
%\par
%The most important case for application of this bound is when the
%initial state $\ket{\phi}$ is maximally entangled: for then
%the corresponding $\Delta_0$ is $0$, so the communication cost is
%essentially lower bounded by $\Delta_\epsilon$ of the final state.

\section{Entanglement concentration and dilution}
\label{sec:dilution}
In~\cite{bbps} it was shown that, using only local operations, Alice and 
Bob can convert a state $\ket{\psi_{AB}}^{\otimes n}$ to a high fidelity 
approximation of $\ket{\phi_2^+}^{\otimes nE(\psi)-O\left(\sqrt{n}\right)}$.
We reproduce the argument here, as the relevant concepts are used again
in the dilution protocol and our lower bound.
\par
Diagonalise $\rho_A=\tr_B\ketbra{\psi}=\sum_{i=1}^d r_i \ketbra{e_i}$.
For a distribution $P$ on $\{1,\ldots,d\}$
we can introduce the \emph{type classes} of sequences $i^n=i_1\ldots i_n$:
$$\fset{T}_P^n:=\left\{ i^n : \forall i\ N(i|i^n)=nP(i) \right\},$$
where $N(i|i^n)$ counts the number of occurences of $i$ in $i^n$.
The number of non--empty type classes is
$\binom{n+d-1}{d-1} \leq (n+1)^{d}$, and the corresponding
$P$ are called $n$--\emph{types}.
\par
For $\delta>0$ we have the set of \emph{typical sequences}
$$\fset{T}_{r,\delta}^n:=\bigcup\left\{ \fset{T}_P^n: P\text{ s.t. }\forall i\
                            |P_i-r_i| \leq \frac{\delta\sqrt{r_i(1-r_i)}}{\sqrt{n}} \right\}.$$
Standard facts about these concepts are to be found in~\cite{wolfowitz}
(see also~\cite{csiszar:koerner}):
\begin{align}
  \label{eq:typical:prob}
  r^{\otimes n}\left(\fset{T}_{r,\delta}^n\right)    &\geq 1-\frac{d}{\delta^2}, \\
  \label{eq:type:prob}
  \forall i^n\in\fset{T}_P^n\quad r^{\otimes n}(i^n) &=\exp\bigl(-n( D(P\|r)+H(P))\bigr),
\end{align}
with the relative entropy (or entropy divergence)
$D(P\|r)=\sum_i P_i\log\frac{P_i}{r_i}$.
Furthermore,
\begin{align}
  \label{eq:typical:card}
  \left| \fset{T}_{r,\delta}^n \right| &\leq \exp\bigl( nH(r)+Kd\delta\sqrt{n} \bigr), \\
  \label{eq:type:cardi}
  \left| \fset{T}_P^n \right|          &\leq \exp\bigl( nH(P) \bigr),                  \\
  \label{eq:type:card}
  \left| \fset{T}_P^n \right|          &\geq (n+1)^{-d}\exp\bigl( nH(P) \bigr),        \\
  \label{eq:typical:type:card}
  \left| \fset{T}_P^n \right|          &\geq \exp\bigl( nH(r)-Kd\delta\sqrt{n} \bigr)
  \quad\text{if }P\text{ typical},
\end{align}
for an absolute constant $K>0$. These sets allow for the definition of 
corresponding
projectors $\Pi_P^n:=\sum_{i^n\in\fset{T}_P^n} \ketbra{e_{i^n}}$,
and similarly $\Pi_{\rho,\delta}^n$, with probability and trace relations
identical to eqs.~(\ref{eq:typical:prob}--\ref{eq:typical:type:card}).
Note that $H(r)=S(\rho)$, by definition.
\par
The concentration protocol only requires Alice and Bob to each independently
perform the projective measurement $(\Pi_P^n)_{P\ n\text{--type}}$. 
(Without loss of generality $\ket{\psi}$ is in Schmidt 
diagonal form, and the bases with respect to which
the projectors are defined are \emph{identical} eigenbases of
the reduced states.)
The result $P$ will be the same for Alice and Bob, and
by eq.~(\ref{eq:typical:prob}) it will be typical with probability
$\geq 1-\epsilon$ (choosing $\delta$ large enough).  Moreover,  by
eq.~(\ref{eq:typical:type:card}) the resulting states $\ket{\phi_P}$
are maximally entangled states of Schmidt rank
$\geq \exp\bigl( nH(r)-Kd\delta\sqrt{n} \bigr)$. Local measurements,
corresponding to a partition of $\fset{T}_P^n$ into blocks of
size $2^m$ (and a remainder of smaller size), for
$m=\bigl\lfloor nH(r)-Kd\delta\sqrt{n}+\log\epsilon \bigr\rfloor$,
project this further down to a state isomorphic to
$\ket{\phi_2^+}^{\otimes m}$, with probability $1-\epsilon$.
This shows that $\ket{\psi}^{\otimes n}$ can be converted 
by local operations into $m$ EPR pairs, with fidelity $1-2\epsilon$,
establishing that asymptotically $\ket{\psi}$ is worth $E(\psi)$
EPR pairs.
\par
In the same work it was demonstrated that the reverse is true as well:
using LOCC, $\ket{\phi_2^+}^{\otimes nE(\psi)+O\left(\sqrt{n}\right)}$
can be converted to a high fidelity approximation of
$\ket{\psi_{AB}}^{\otimes n}$.
\par
Alice simply prepares the state
$\Pi_{\rho,\delta}^n\ket{\psi}^{\otimes n}$ (properly normalized)
locally.
By eq.~(\ref{eq:typical:card}) it has Schmidt rank
$\leq \exp\bigl( nH(r)+Kd\delta\sqrt{n} \bigr)$, enabling Alice to
teleport~\cite{teleportation}
the half intended for Bob using $nH(r)+Kd\delta\sqrt{n}$ EPR pairs.
\par
Note that this method requires communication of
$2nE(\psi)+O\left(\sqrt{n}\right)$ classical bits from Alice to Bob,
which is of the order of the entanglement manipulated.
Whether this amount can be reduced is, therefore, a legitimate and 
interesting question.
In~\cite{lo:popescu:sqrt} it was shown that, indeed,
communication of $O\left(\sqrt{n}\right)$ classical bits are sufficient,
by the following method:
\par
They demonstrated that there exists a state $\ket{\chi}$ entangling $O\left(\sqrt{n}\right)$
qubits, and local unitaries $U_A$ and $U_B$ such that
\begin{equation}
  \label{eq:chi:fidelity}
  F\left( (U_A\otimes U_B)\ket{\psi}^{\otimes n},
          \ket{\phi_2^+}^{\otimes nE-O\left(\sqrt{n}\right)}\otimes\ket{\chi} \right)
                                                                        \geq 1-\epsilon.
\end{equation}
This state arises naturally by looking at what was done in the concentration
procedure above, in a reversible setting.  Applying the same dilution
procedure as before but to the smaller state $\ket{\chi}$, that is, 
local preparation by Alice followed by teleportation of Bob's share, 
then only consumes  $O\left(\sqrt{n}\right)$ 
ebits and twice that amount of classical communication
(as Lo~\cite{lo:ccc} has shown this factor can be reduced to $1$, i.e., a state of
Schmidt rank $d$ can be prepared using $\log d $ bits of entanglement and communicating
$\log d$ classical bits).
\par
Let us now apply our main result to show that any protocol to create $\ket{\psi}^{\otimes n}$
up to fidelity $1-\epsilon$ from EPR pairs must use
$\Omega\left(\sqrt{n}\right)$ bits of communication:
\par
Noting first that EPR pairs have $\Delta_0=0$, we have only to lower bound
$\Delta_\delta\left(\psi_A^{\otimes n}\right)$ in order to make use of 
theorem~\ref{thm:bound}. This
we do by using eq.~(\ref{eq:piecemeal}). First, we show that
$$S_{\infty,\delta}\left(\psi_A^{\otimes n}\right) 
                  \leq nE(\psi)-D(\epsilon)\sqrt{n}+o\left(\sqrt{n}\right),$$
with a constant $D(\epsilon)>0$ (for $\delta=\sqrt[8]{4\epsilon}<1/2$):
\par
Observe that $S_{\infty,\delta}$ is particularly easy to understand;
it is the negative logarithm of the largest eigenvalue such that the sum of the
eigenvalues exceeding this one is bounded by $\delta$.
\par
Define the independent and identically distributed (i.i.d.) random variables
$X_j$, $j=1,\ldots,n$, by letting
$$\Pr\{ X_j=-\log r_i \} = r_i,$$
where $(r_i)$ are the Schmidt coefficients of $\ket{\psi}$.
Note that their expectation $\E X_j$ equals $E(\psi)$, and that they are
nonconstant, unless $\ket{\psi}$ is maximally entangled, so that
the variance $\sigma^2$ is nonzero.
\par
Hence we can apply the central limit theorem:
$$\Pr\left\{ \sum_{j=1}^n X_j \leq nE(\psi)+x\sigma\sqrt{n} \right\}
                 \longrightarrow \frac{1}{\sqrt{2\pi}}\int_{-\infty}^x \!\! e^{-t^2/2}{\rm d}t.$$
This implies that the sum of the largest eigenvalues, from
$\exp\bigl( -nE(\psi)+D(\epsilon)\sqrt{n}+o(\sqrt{n}) \bigr)$ up
(including multiplicities), is bounded from below
by $\delta$, and our claim follows.
\par
Next, we lower bound $S_{0,\delta}\left(\psi_A^{\otimes n}\right)$.
An optimal set $J$ in the definition, eq.~(\ref{eq:S:0:epsilon}),
must consist of the indices of the $|J|$ largest eigenvalues
such that their sum is barely above $1-\epsilon$.
\par
Once more invoking the central limit theorem, the sum of the smallest
eigenvalues (including multiplicities) of $\psi_A^{\otimes n}$ up to
$\exp\bigl( -nE(\psi)-D(\epsilon)\sqrt{n}+o(\sqrt{n}) \bigr)$
is at least $\delta$.
\par
We exhibit now a large type class inside the set corresponding to larger
eigenvalues, which by the preceding is a subset of $J$:
there exists an $n$--type $P$ such that
$|P_i-r_i|\leq 1/n$, for all $i$. This entails that for $i^n\in{\cal T}_P^n$
\begin{equation*}\begin{split}
  \log r^{\otimes n}(i^n) &= n \sum_i P_i \log r_i                            \\
                          &= n \sum_i \left(r_i\pm\frac{1}{n}\right) \log r_i \\
                          &= -nH(r) \pm \sum_i |\log r_i|                      \\
                          &= -nE(\psi) \pm C.
\end{split}\end{equation*}
Thus, ${\cal T}_P^n\subset J$ as soon as $D(\epsilon)>0$ and $n$ is large enough.
\par
On the other hand, because $\|P-r\|_1\leq d/n$, we have
(using a well--known estimate for Shannon entropies,
see~\cite{csiszar:koerner}) that
$$| H(P) - H(r) | \leq \frac{d}{n}\log n,$$
and we conclude, by eq.~(\ref{eq:type:card}), that
\begin{equation*}\begin{split}
  |J| &\geq |{\cal T}_P^n| \\
      &\geq (n+1)^{-d}\exp\bigl( nH(P) \bigr) \\
      &\geq \exp\bigl( nH(r)-d\log n-d\log(n+1) \bigr).
\end{split}\end{equation*}
It follows that $S_{0,\delta}(\psi_A^{\otimes n}) \geq nE(\psi)-O(\log n)$.
\par
Combining the estimates of $S_{0,\delta}$ and $S_{\infty,\delta}$,
we obtain
\begin{thm}
  \label{thm:dilution:sqrt}
  For every bipartite pure state $\ket{\psi_{AB}}$ that is neither separable
  nor maximally entangled
  and every sufficiently small $\epsilon$ there exists a positive constant $D(\epsilon)$
  such that the communication cost of creating $\ket{\psi}^{\otimes n}$ up
  to fidelity $1-\epsilon$ from EPR pairs
  is at least $C \geq D(\epsilon)\sqrt{n}-o\left(\sqrt{n}\right)$.
  \qed
\end{thm}
\begin{rem}
  \label{rem:classical:analogue}
  Recently, secret shared randomness has been proposed as a ``classical analogue of
  entanglement''~\cite{collins:popescu}, partly to increase intuition on entanglement
  transformations, and partly to be able to distinguish the quantum effects 
  of entanglement from those that are statistically explainable.
  \par
  Specifically, pure state entanglement was parallelled to classical
  \emph{perfect} correlation: Alice and Bob share a joint random variable
  $(X,Y)$, where $X$ belongs to Alice, $Y$ to Bob and $X=Y$ with 
  probability $1$. Entanglement transformations by LOCC
  have their analogue in transformations of these random variables
  by local (classical) actions and \emph{public discussion},
  which can be listened to by an eavesdropper. The analogue of EPR
  pairs are shared random bits: $\Pr\{X=Y=0\}=\Pr\{X=Y=1\}=1/2$.
  \par
  Now it is an easy result of the theory of shared randomness
  (see~\cite{ahlswede:csiszar} for definitions)
  that in an i.i.d. setting $(X,Y)$ can be asymptotically
  converted into the Shannon entropy $H(X)$ of $X$ many
  shared secret bits and, inversely, this amount of shared
  randomness can be used to generate $(X,Y)$: more precisely,
  both transformations can be performed with asymptotically vanishing 
  total variational distance of the distributions. These operations
  are the classical analogues of entanglement concentration and dilution.
  \par
  What is remarkable is that in this setting both the concentration and  
  dilution processes require \emph{no public discussion whatsoever}. 
  Thus, our $\Omega\left(\sqrt{n}\right)$
  lower bound is a purely \emph{quantum} phenomenon that has no
  counterpart in the ``classical analogue''.
\end{rem}

\section{States with large\protect\\ communication cost}
\label{sec:large:cost}
In~\cite{embezzle}, the states
$$\ket{\mu(n)}=\frac{1}{\sqrt{H_n}} \sum_{i=1}^n \frac{1}{\sqrt{i}}\ket{i}\ket{i},$$
with the harmonic sum $H_n=\sum_{i=1}^n \frac{1}{i}$, were introduced to show that
the concept of ``approximate pure state transformations with unlimited
catalysis'' allows \emph{any} state transformation (this was dubbed
``embezzling entanglement'' in~\cite{embezzle}). In particular
it was shown that for every pure state $\ket{\phi}$ of Schmidt rank
$m$ there are local isometries $U_A$ and $U_B$ such that
$$F\bigl(\ket{\mu(n)}\otimes\ket{\phi},(U_A\otimes U_B)\ket{\mu(n)}\bigr)
                                                     \geq 1-\frac{\log m}{\log n}.$$
It is straightforward to verify that the entanglement of $\ket{\mu(n)}$ 
is asymptotically $\frac{1}{2}\log n$, and we shall demonstrate here that
the communication cost to produce it from EPR pairs is of the same order:
\par
Theorem~\ref{thm:bound} asks us to lower bound $\Delta_\delta$ of Alice's
reduced state
$$\rho_A=\frac{1}{H_n}\sum_{i=1}^n \frac{1}{i}\ketbra{i},$$
which we do using eq.~(\ref{eq:piecemeal}):
\begin{align}
  \label{eq:mu:rank}
  S_{0,\delta}(\rho_A)      &= \log\min\left\{ k:\sum_{i=k+1}^n \frac{1}{iH_n}   \leq \delta \right\}, \\
  \label{eq:mu:eigen}
  S_{\infty,\delta}(\rho_A) &= \log H_n
                             +\log\max\left\{ k:\sum_{i=1}^{k-1} \frac{1}{iH_n} \leq \delta \right\}.
\end{align}
Now, asymptotically $(\log n)-1\leq H_n \leq \log(n+1)$, and eqs.~(\ref{eq:mu:rank})
and~(\ref{eq:mu:eigen}) allow us to estimate
\begin{equation}
  \label{eqn:largecost}
  \Delta_\delta(\rho_A) \geq \bigl((1-2\delta)\log n\bigr)-4-\log\log(n+1),
\end{equation}
resulting in a lower bound
$$C \geq \left(\frac{1}{2}-\sqrt[8]{4\epsilon}\right)\log n - O(\log\log n)$$
for the communication cost to create $\ket{\mu(n)}$ up to fidelity $1-\epsilon$
from EPR pairs. In fact, the classical communication cost is, by
corollary~\ref{cor:classical}, lower bounded by $\bigl(1-o(1)\bigr)\log n$,
asymptotically matching the upper bound $\log n$ from Lo's earlier mentioned
state preparation method in~\cite{lo:ccc}.
\par\medskip
Other states with entanglement being of the same order as the communication
necessary to create them are the $\ket{\chi}$ of eq.~(\ref{eq:chi:fidelity}):
their entanglement is at most $O(\sqrt{n})$ while theorem~\ref{thm:dilution:sqrt}
implies a lower bound of $\Omega(\sqrt{n})$ on the communication resources.

\section{Conclusion}
\label{sec:conclusion}
We have exhibited the first quantitative lower bound on the communication cost
of general entanglement transformations. It is good enough to prove that the Lo/Popescu
protocol of entanglement dilution is within a constant factor of being optimal,
requiring $\Theta\left(\sqrt{n}\right)$ bits of communication.
Also, it can be used to show that there exist states whose communication cost
for creation from EPR pairs is of the same order as their entanglement, making
local preparation and teleportation essentially the optimal strategy.
\par
It is unknown to us how tight our lower bound can be made or if there is an 
upper bound involving similar quantities, so we leave these questions open 
for future research.
On a different note, it has repeatedly been speculated (such as 
in~\cite{lo:popescu:sqrt}) that the classical communication cost is 
related to the \emph{loss of entanglement} in a transformation.
Observe that this seems to fit perfectly
for concentration and dilution, and it might be that in an appropriate model
the entanglement loss in a pure state transformation provides an upper
bound on the minimal communication cost required to perform it.
\par
Other applications may include the study of quantum communication complexity,
where a technique for lower bounding the communication 
exists~\cite{cdnt,astvw,comm:compl} that requires estimation of 
the communication cost
of certain pure state entanglement transformations. In the cited works
this cost was lower bounded by observing that some measure of entanglement has
increased. Our method could be useful as it gives nontrivial lower bounds even
when the entanglement remains constant or decreases, and continues to
be effective in the presence of unlimited numbers of EPR pairs.
\par\medskip
After the present paper was finished, the independent work of
Harrow and Lo~\cite{harrow:lo} came to our attention,
which proves the $\Omega(\sqrt{n})$ lower bound
on entanglement dilution by a different method (though there are
similarities) that simultaneously provides a lower bound on 
the entanglement loss.

\acknowledgments
We thank Wim van Dam for his suggestion to also consider general 
R\'{e}nyi entropies, and Karol and Michal Horodecki for posing the
problem solved in remark~\ref{rem:epr2epr}. We want to thank Aram Harrow
and Hoi--Kwong Lo for making their draft of~\cite{harrow:lo} available
to us and for stimulating discussions.
\par
P.H.~was supported by US National Science Foundation grant no.~EIA--0086038
and a Sherman Fairchild Fellowship.
A.W.~is supported by the U.K.~Engineering an Physical Sciences Research Council.
This work was carried out during the second author's visit to
the Institute of Quantum Information, Caltech, in January 2002.

% bibliographie

%\bibliographystyle{revtex}

\end{document}